# Exploiting the Electrothermal Timescale in PrMnO$_3$ RRAM for a compact, clock-less neuron exhibiting biological spiking patterns

Omkar Phadke[1], Jayatika Sakhuja[1], Vivek Saraswat[1], Udayan Ganguly[1]

[1]Department of Electrical Engineering, IIT Bombay, Mumbai, India.
Email: udayan@ee.iitb.ac.in

## Abstract

Spiking Neural Networks (SNNs) are gaining widespread momentum in the field of neuromorphic computing. These network systems integrated with neurons and synapses provide computational efficiency by mimicking the human brain. It is desired to incorporate the biological neuronal dynamics, including complex spiking patterns which represent diverse brain activities within the neural networks. Earlier hardware realization of neurons was (1) area intensive because of large capacitors in the circuit design, (2) neuronal spiking patterns were demonstrated with clocked neurons at the device level. To achieve more realistic biological neuron spiking behavior, emerging memristive devices are considered promising alternatives. In this paper, we propose, PrMnO3(PMO) -RRAM device-based neuron. The voltage-controlled electrothermal timescales of the compact PMO RRAM device replace the electrical timescales of charging a large capacitor. The electrothermal timescale is used to implement an integration block with multiple voltage-controlled timescales coupled with a refractory block to generate biological neuronal dynamics. Here, first, a Verilog-A implementation of the thermal device model is demonstrated, which captures the current-temperature dynamics of the PMO device. Second, a driving circuitry is designed to mimic different spiking patterns of cortical neurons, including Intrinsic bursting (IB) and Chattering (CH). Third, a neuron circuit model is simulated, which includes the PMO RRAM device model and the driving circuitry to demonstrate the asynchronous neuron behavior. Finally, a hardware-software hybrid analysis is done in which the PMO RRAM device is experimentally characterized to mimic neuron spiking dynamics. The work presents a realizable and more biologically comparable hardware-efficient solution for large-scale SNNs.

## 1. Introduction

A Spiking Neural Network (SNN) mimics the human brain to perform complex computations. The ability to perform tasks parallelly, along with low power consumption, gives SNN an edge over traditional von Neumann architecture, which is limited by the von Neumann bottleneck. The human brain consists of two main components; a neuron, which fires according to the input signal, and a synapse, which interconnects two neurons. Whenever a neuron receives a stimulus, it integrates the input signal to its membrane potential. Once the membrane potential reaches a threshold, it falls back down to low levels, exhibiting a spike. The spike timing of the neuron and the strength of the synapse are crucial for information processing. As illustrated in Fig.1, a neuron can exhibit different spiking patterns, and each spiking pattern plays a unique role in the functioning of the brain [1]. For example, neurons that exhibit 'Tonic Bursting' are thought to be responsible for the 'Gamma Wave Oscillations' in the brain.

Some neurons are also used for concentration gradient computations in chemotaxis in C. elegans, which is used to translate from biology to algorithms [2]. Since the biological neurons exhibit a rich variety of spiking patterns, we need artificial neurons which mimic these spiking patterns to build SNNs which can perform a wider variety of tasks (Fig.1). Further, the refractory period occurs after spiking when the neuron is quiescent and does not effectively integrate inputs. The tuning of the refractory period is another critical element of the neuron to control the dynamics of brain waves [3] and navigational circuits [4]. Thus, the neuron has various timescales – both during spiking and refractory phases.

Previously, many mathematical models of the neuron have demonstrated different spiking patterns of the cortical neuron. A popular model has been the Izhikevich Model, which is biologically plausible and computationally efficient [5]. While such models help study large-scale SNNs in a simulation environment and providing us insights into brain functioning, they are not useful for hardware implementation of SNNs. For hardware realization of SNNs, we need devices that mimic different spiking patterns of the neuron while being area efficient and consuming less power.

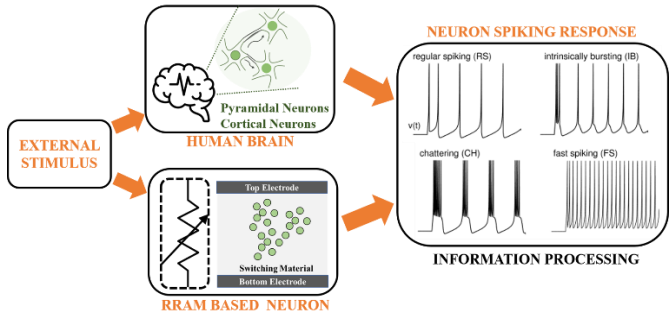

**Figure 1** Motivation figure. Implementation of diverse neuronal spiking patterns observed in human brain. through RRAM devices will be computationally efficient for neural networks.

Various device-based neurons have been demonstrated [6]-[11]. Here, the 'integration' operation is implemented by charging the external capacitor. The size of the capacitor and the input determine the spiking frequency, and to show different integration timescales, the area of the capacitor needs to be changed. Also, these neurons consume a large area, making the large-scale implementation of SNNs challenging especially when realizing longer real-world signal timescales (e.g., speech). Different spiking patterns have also been implemented using digital circuitry [12]. The implementation can demonstrate a wide range of spiking patterns while being biologically plausible. However, the circuitry of a single neuron is complex as it involves multiple buffers, counters, multiplexers, and pipelines (which consist of adders), and the implementation is clocked. It shows an implementation for a small SNN, but for bigger SNNs, the resulting digital circuitry will be large and complex.

Recently, Resistive Random-Access Memory (RRAM) based implementation of neurons was demonstrated, which utilized the internal timescales of the RRAM device (either gradual resistive switching or electrothermal timescale) to perform the integration operation, which eliminated the need of using an external capacitor [13]-[14]. Specifically, a $Pr_{1-x}Ca_xMnO_3$ (PCMO) based RRAM device was proposed to implement an Integrate and Fire (IF) neuron [13]. The paper also demonstrated different spiking patterns, such as Intrinsically Bursting (IB), and Chattering (CH) patterns. Also, the neuron utilized the internal timescale of resistive switching to perform integration operation. The PCMO RRAM can show different conductance levels and exhibits different spiking frequencies for these different conductance levels. These two features eliminated the need for an external capacitor. The major disadvantages of the PCMO RRAM neuron implementation were that the neuron did not exhibit a 'Leaky' behavior, which is integral to the functioning of a biological neuron, and it utilizes a clock to operate, whereas human brains do not use a clock [15].

A $PrMnO_3$ (PMO) RRAM-based Leaky IF (LIF) neuron was demonstrated experimentally [14]. PMO-based RRAM devices are non-filamentary and highly scalable, making them attractive for compact neurons. Here, the internal electrothermal timescales in PMO material were used to perform integration, eliminating the use of an external capacitor. The PMO RRAM-based neuron was asynchronous and demonstrated only a single spiking pattern.

In this paper, we demonstrate an asynchronous, capacitor-free, PMO RRAM-based neuron that can exhibit different spiking patterns of a cortical neuron. The neuron utilizes the multiple voltage-controlled electrothermal timescales of PMO RRAM to construct an integration block coupled to a refractory block to enable a compact capacitor-free timescale control. We propose a simulation setup consisting of a physics-based Verilog-A model of the RRAM, and a behaviorally modeled driving circuit, to model and predict different spiking patterns. The voltage across the device from the simulation setup is extracted, approximated, and applied across the PMO RRAM to experimentally demonstrate the different biological spiking patterns. The work provides a strategy to investigate different biological spiking patterns in simulations as well as experimentally in area-efficient memristor-based synaptic arrays.

## 2. Device Details

The stack of PMO-based RRAM is as shown in Fig.2a. The Silicon (Si) substrate is used for RRAM fabrication. $SiO_2$ is thermally grown on the Si wafer followed by deposition of a bilayer of Titanium (Ti) followed by Platinum (Pt) deposited through DC sputtering. A blanket layer of PMO film (60 nm) is then deposited using RF sputtering at room temperature. The PMO film is annealed in $O_2$ ambient at 750°C for 30 seconds. Finally, the top metal, tungsten (W), contact pads are patterned through UV lithography and metal lift-off process.

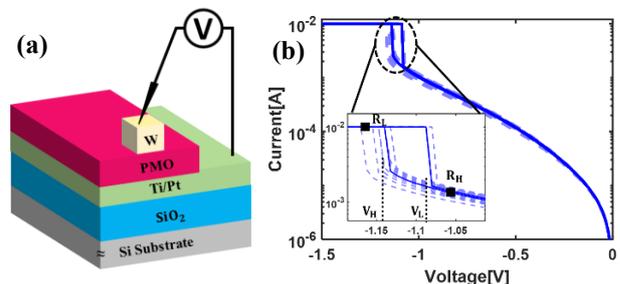

**Figure 2** (a) Device Schematic of fabricated PMO based RRAM (b) The volatile DC-IV characteristics of PMO RRAM device with C2C variation. The dotted blue lines show data for different cycles. The solid blue line is the mean value. The inset shows the enlarged version of the volatile hysteresis loop.

The resulting device structure is a metal-insulator-metal (MIM) structure, with reactive metal W and noble metal Pt as the top and bottom electrodes, as shown in Fig. 2a. If a negative bias is applied to the reactive electrode, the device goes to a Low Resistance State (LRS). If a positive bias is applied to the reactive electrode, the device goes to a High Resistance State (HRS). The non-volatile resistance change occurs due to modulation of trap density at the reactive electrode interface. The current through the device follows the Space Charge Limited Current (SCLC) mechanism [16]-[17]. When the device is in LRS state, and a negative voltage sweep is applied, the device exhibits a volatile-resistance change. The DC IV characteristics of volatile switching are as shown in Fig. 2b. For the experiments demonstrated in the paper, the device is in LRS. The volatile switching shown by the device occurs due to the electrothermal mechanism. As voltage is applied, a current start to flow through the device. At higher currents the device temperature increases, owing to Joule heating, which in turn increases the current through the device. The generated heat is trapped inside the device due to the low thermal conductivity of the PMO material (1.48Wcm-1K-1)[18]. At a certain threshold voltage, the interplay between the current and temperature results in positive feedback, which leads to a sharp shoot-up in current.

## 3. Electrothermal RRAM Device Model

To capture the device's current-temperature dynamics, an electrothermal model is demonstrated. The current is calculated using the analytical model shown in equations (1-3), and the temperature is calculated using equation (4) [16].

$$I = I_{Ohmic} + I_{SCLC} \tag{1}$$

$$I_{Ohmic} = qA\mu N_v \left(\frac{T}{T_{amb}}\right)^{3/2} e^{\left(-\frac{q\phi_B}{kT}\right)} \left(\frac{V}{L}\right) \tag{2}$$

$$I_{SCLC} = A\mu\epsilon_o\epsilon_{PMO} \left(\frac{N_v}{N_T}\right) \left(\frac{T}{T_{amb}}\right)^{3/2} e^{\left(-\frac{qE_{trap}}{kT}\right)} \left(\frac{V^2}{L^3}\right) \tag{3}$$

$$\frac{T - T_{amb}}{R_{th}} + C_{th}\frac{dT}{dt} = IV \tag{4}$$

The equations are modeled in Verilog-A and simulated in Cadence Virtuoso. The flowchart for the model is as shown in Fig. 3. The model is initialized by setting the temperature as ambient temperature (300K) and applying voltage bias. Using equations (1) to (4), temperature and current are solved self-consistently.

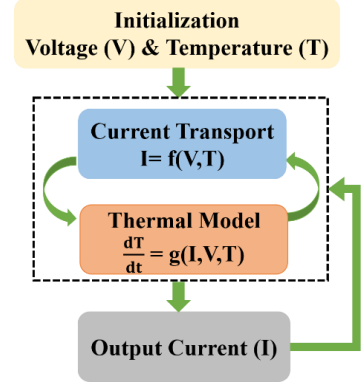

**Figure 3** Electrothermal Device Model Flowchart, illustrating the positive thermal feedback to capture self-heating dynamics of the device.

The equivalent thermal coefficients are $R_{th}$ (thermal resistance) and $C_{th}$ (thermal capacitance). Here, the temperature $T$ is an "effective" temperature of the device. Fig. 4a and 4b show the transient currents for experiments and simulations for different applied voltages. The spike time of the model is calibrated to match those of experiments, and the results are shown in Fig. 4c. The parameters used in simulations are mentioned in Table 1.

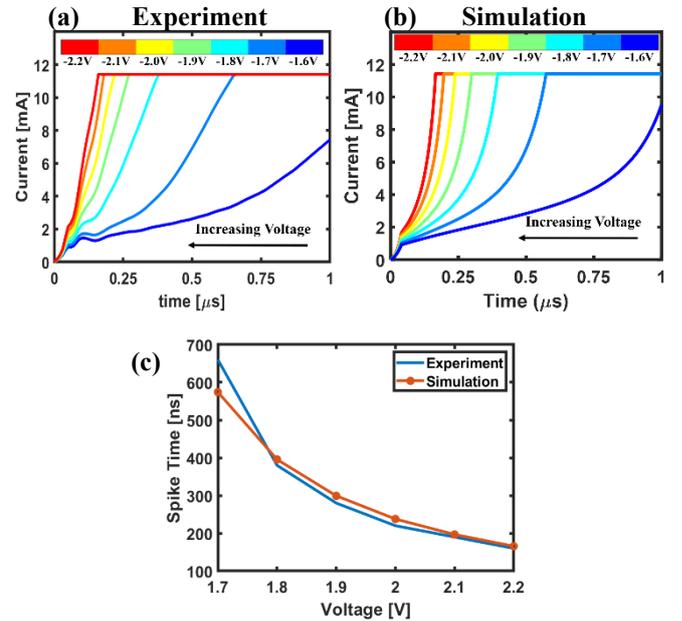

**Figure 4** Transient current response of RRAM for various applied voltage pulses. (a) Experiments (b) Simulations (c) Calibration of Spike Time. Increasing voltage reduces the time required to reach the current compliance (Spike Time) is attributed to self-heating within device.

| Model | Symbol | Quantity | Value |
|---|---|---|---|
| Analytical Model | μ | Mobility | 17.5 cm$^2$/Vs |
| | φ$_B$ | Barrier Height | 0.3151 eV |
| | ε$_{PMO}$ | Dielectric Constant | 30 |
| | N$_v$ | Density of States | 8.16×10$^{19}$ cm$^{-3}$ |
| | E$_{trap}$ | Trap Level | 0.06 eV |
| | N$_T$ | Trap Density | 3.15×10$^{21}$ cm$^{-3}$ |
| Device | L | Length | 65 nm |
| | A | Area | 10×10 μm$^2$ |
| Thermal Model | T$_{amb}$ | Ambient Temperature | 300 K |
| | R$_{th}$ | Thermal Resistance | 3×10$^4$ K/W |
| | C$_{th}$ | Thermal Capacitance | 3.25 pJ/K |

**Table 1:** Parameters Used in the Model

## 4. Input Voltage Dependent Electrothermal Timescale Control

As shown in the flowchart Fig. 3, current and temperature are dependent quantities. Therefore, with an increase in current, temperature increases, which further increases the current. The positive feedback between current and temperature results in a current shoot-up. A constant voltage pulse of different magnitudes is applied across the device, and the current through the device is observed (Fig. 4). The time for the current to shoot up and reach the compliance depends upon the applied voltage. Thus, a higher voltage will trigger a faster shoot-up, and a lower voltage will slow down the current shoot-up. The current-temperature time dynamic is used as an integration timescale in the proposed PMO RRAM-based neuron.

## 5. Circuit Implementation:

### 5.1 Exhibiting Different Spiking Patterns

Biological neurons are capable of exhibiting different spiking patterns. To mimic that behavior in hardware using RRAM, the following circuit concept and operation are proposed. A neuron integrates the input with a timescale to raise its membrane potential, and a spike is issued when a threshold is reached. The spike patterns are characterized by the positioning of spikes or spike timings for constant input. A neuron based on RRAM can generate spike patterns by using the electrothermal timescale for integration controlled by the input voltage, which is slowly time-varying input from synapses. Complex biological spiking patterns can be generated by modulating the input voltage applied either directly or with an additional resistive drop across RRAM based on an internal binary state variable of the neuron to modulate the integration timescales dynamically. Further, the refractory period control is enabled by coupling another RRAM block to the integration block – whose electrothermal timescale is voltage-controlled.

To enable the above operation, the driver circuit shown in Fig. 5 is proposed. The circuit is capable of showing different spiking patterns. The working of each component is explained below. RRAM is referred to as R$_D$.

#### a. Resetting Switch (S1, S2)

The switches S1, S2 are used to connect or disconnect the voltage source from the RRAM. If the input to the switch is 0, the switch is closed, connecting the input voltage to the RRAM. If the input to the switch is 1, the switch is opened, disconnecting the input voltage from RRAM.

#### b. Series Resistor (R$_S$)

A series resistor (R$_S$) is connected in series with the RRAM. The voltage drop across R$_S$ (V$_A$) is linearly dependent on the current through RRAM. If the current exceeds a threshold, so does V$_A$. The voltage V$_A$ is used to detect spikes.

#### c. 4-bit Register (Register 1)

A register is implemented, which stores a sequence of 1s and 0s according to the desired spiking pattern. For example, the register would initially store 1110 bits for the Chattering (CH) pattern, which will lead to 3 spikes with a smaller spike time and 1 with a larger spike time. It requires a trigger signal to perform the left-shift operation, which it receives from the voltage V$_A$ once the neuron has fired.

#### d. Control Resistor (R$_C$)

Control Resistor has two states; it either acts as a short circuit (SC) or as a resistor. When R$_C$ acts as a SC, the input voltage drops across the (S1+R$_{D1}$+R$_{S1}$) network. When R$_C$ acts as a resistor, a reduced voltage drops across the (S1+R$_{D1}$+R$_{S1}$ network). The MSB of the Register 1 controls the state of the R$_C$; if MSB = 1, R$_C$ acts as a SC, and if MSB = 0, R$_C$ acts as a resistor.

#### e. Refractory Block

In a biological neuron, after a neuron has fired, it does not respond to the external stimulus for a period called the 'Refractory Period.' On a circuit level, control over this refractory period is desirable. The driving circuit consists of two neurons; one neuron (in the Integration Block) is driven by the input signal, and the second neuron (in the Refractory Block) is driven by a constant voltage signal which controls the refractory period.

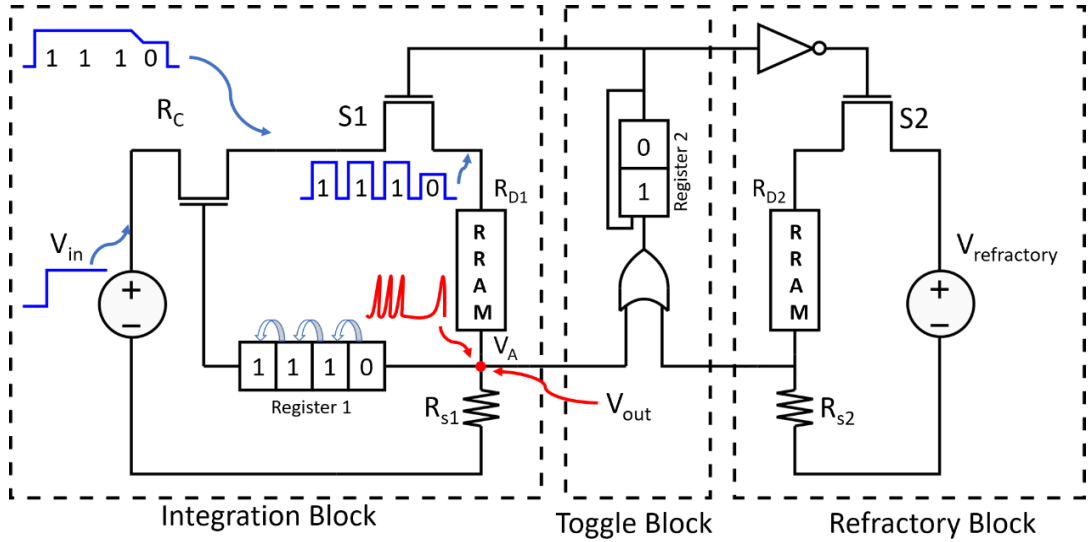

**Figure 5** Simulation Setup of Driving Circuit of the RRAM based Neuron to produce different spiking patterns. The Integration Block receives input stimulus and produce spiking patterns. The refractory period of the spiking patterns is controlled by the Refractory Block.

*f. Toggle Block*

The toggle block ensures that either the Integration Block or the Refractory block stays active at a given time. The toggle block has an OR gate, which detects if either of the neurons has spiked via voltage $V_A$, and a 2-bit register (Register 2) whose MSB controls the switches S1 and S2. Once a neuron has fired, the 2-bit register performs a left shift operation and flips the MSB bit (as only 0 and 1 is stored, and MSB is connected back to LSB). This disconnects the neuron which has fired from the voltage source and reconnects the other neuron to its voltage supply. This operation of switching between 2 blocks is repeated as long as the neuron spikes.

In Fig. 6, a state diagram and a timing diagram is demonstrated, which explains the interplay between the input neuron and the refractory neuron. The state diagram doesn't include the 4-bit register and the control resistor for simplicity. Either the Neuron N1 or N2 remains ON at any given time. The input to the N2 neuron is chosen such that it will always elicit a spike. Once N1 fires, N1 turns OFF and turns ON N2, and vice versa. After N1 has fired, it will stay OFF for a duration equal to the spike time of the N2 neuron, achieving a refractory period. During this duration, N1 will not perform any integration operation. In Fig. 7, a flowchart is demonstrated which explains the working of the neuron to exhibit different spiking patterns. The flow chart doesn't include the refractory block for simplicity. Initially, the input pulse voltage is applied, S1 is closed, and $R_C$ acts as a SC. $V_{in}$ is applied to RRAM, and current is computed according to the RRAM electrothermal model. If the voltage at node A ($V_A$) exceeds the threshold voltage ($V_{th}$), registers 1 and 2 detect the event and perform a left shift operation. For Register 1, if MSB = 1, $R_C$ acts as SC, else acts as a resistor. For Register 2, if MSB = 1, S1 opens up, else closes. To achieve different spiking patterns, a different set of bits are stored in the register. All transitions occur asynchronously, without a need for an external global clock.

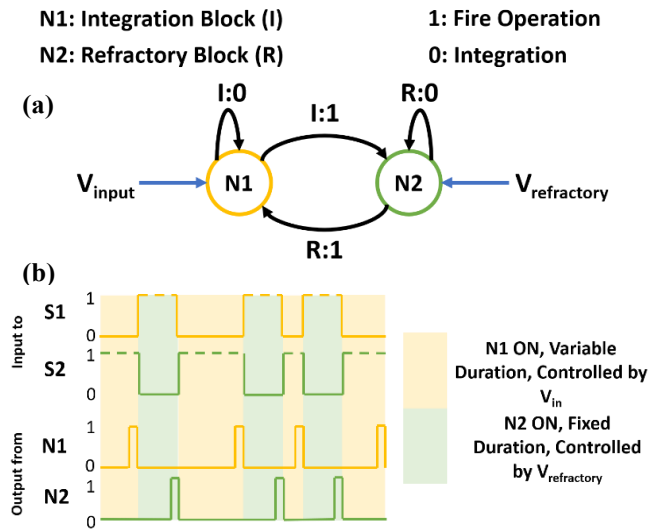

**Figure 6** Interplay between Integration Block and Refractory Block (a) State Diagram. (b) Timing Diagram The state diagram demonstrates the interplay between the two neurons to exhibit a refractory period. Either one of the two neurons will remain ON at any given time.

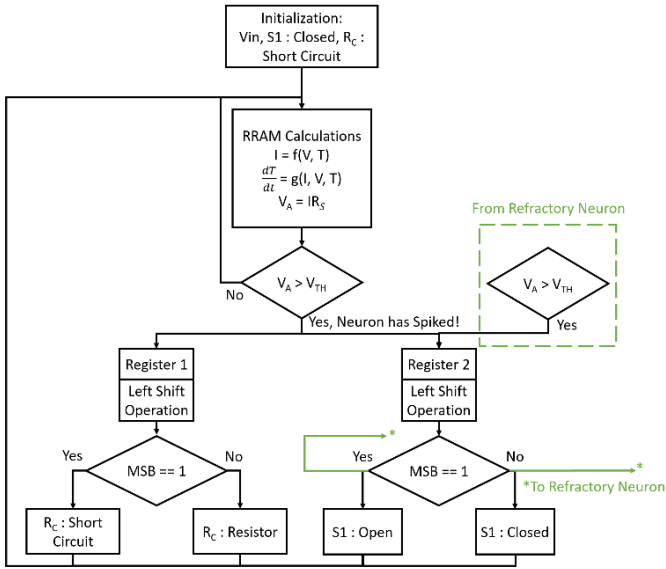

**Figure 7** Flowchart illustrating the control logic implemented of driving circuit to emulate diverse spiking patterns.

## 6. Simulation Results and Discussion:

The components of the driving circuit shown in Fig. 5 are modeled behaviorally in Verilog-A, and the circuit simulations are performed in Cadence Virtuoso. The simulation results are presented and discussed below.

### 6.1 Voltage Controlled Spiking Frequency

Fig. 8 shows the simulation results for $V_{input}$ = -1.6V and $V_{input}$ = -1.8V. As explained in section 4, high input voltage leads to a higher spiking frequency and vice versa.

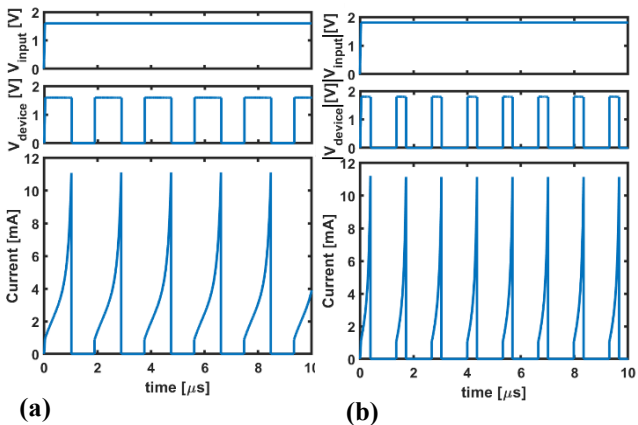

**Figure 8** Simulated Voltage Controlled Spiking Frequency demonstrated. The switch S1 converts the input voltage into discrete voltage pulses, which leads to current spikes. (a) For $V_{input}$ = -1.6V, spiking frequency is 537 KHz, while for (b) $V_{input}$ = -1.8V, the spiking frequency is 754 KHz.

### 6.2 Voltage Controlled Refractory Period

Fig. 9 shows the simulation results for $V_{input}$ = -2.2V, and a refractory period of 200ns and 400ns provided by the refractory neuron. A high input voltage to the refractory neuron ($V_{refractory}$) leads to a smaller refractory period, and a small voltage leads to a longer refractory period. Different voltages modulate the electrothermal timescales of the refractory neuron and hence the spike timing of the refractory neuron. As spike timing of refractory neuron controls the refractory period of the input neuron, $V_{refractory}$ controls the refractory period.

### 6.3 Time Varying Input

The circuit shown in Fig. 5 is used for demonstration, with the control resistor and the register (Register 1) removed. Two sinusoids with frequency = 250 KHz, & 350 KHz, voltage amplitude = -0.7V, and the DC voltage level = -0.7V are superimposed and applied as an input to the circuit. The resulting input signal has two regions with high voltage, two regions with moderate voltage, and two regions with low voltage. Fig. 10 shows the simulation results. The high voltage region leads to a fast and dense spiking, the moderate voltage region leads to a slower and sparsely distributed spiking, while the low voltage region does not issue a single spike.

The output behavior is attributed to the transient Joule heating mechanism in the device. When the voltage is low, the thermal feedback within the device is not established, hence no current shoot-up, resulting in no spikes. As the voltage is increased, the positive feedback builds up, current levels rise, and spiking patterns are observed.

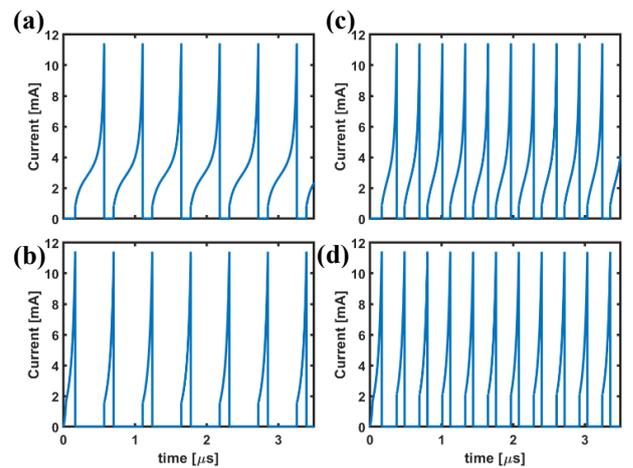

**Figure 9** Simulated Voltage Controlled Refractory Period. The switch S1 converts the input voltage into discrete voltage pulses, which leads to current spikes. (a) Refractory Neuron and (b) Input Neuron for Refractory Period = 400ns, (c) Refractory Neuron and (d) Input Neuron Refractory Period = 200ns.

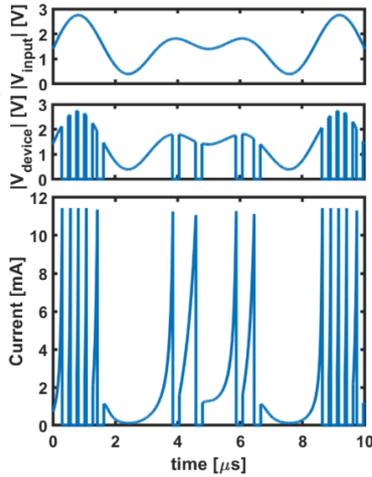

**Figure 10** Simulation Results of Time Varying Input: Input voltage is combination of two-time varying sinusoidal signals of equal magnitude and different frequencies.

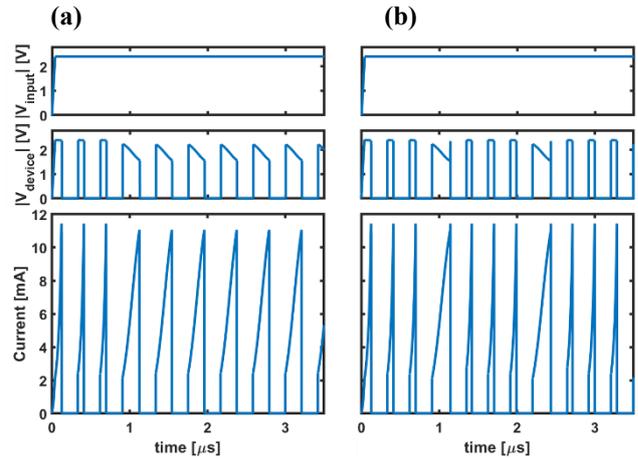

**Figure 11** Simulation Results of Different Spiking Patterns: (a) Intrinsic Bursting (b) Chattering

## *6.4 Different Spiking Patterns*

The circuit shown in Fig. 5 is used. For demonstration, Intrinsic Bursting (IB) and Chattering (CH) patterns are simulated and the results are shown in Fig. 11. In IB, 3 fast spikes are followed by consecutive slow spikes (Fig. 11a). In CH, 3 fast spikes and 1 slow spike are alternated (Fig. 11b).

To achieve CH and IB spiking patterns, bits 1110 are stored in the register. For CH spiking pattern, MSB and LSB are connected to one another, whereas for the IB spiking pattern, MSB and LSB are disconnected from one another. A high voltage is applied at the input terminal. Initial MSB = 1 ensures that the control resistor acts as a switch, and the voltage drops across the $R_{D1}$. Once a spike is issued, the switch S1 opens up, disconnecting $R_{D1}$ from the input voltage source, and voltage $V_A$ triggers the Register 1 to left shifts its contents. Once the spike is over, the register holds 1101 (for CH) and 1100 (for IB). The next two spikes will elicit a similar current response of faster spikes. At the end of the 3 spikes, the contents of the register would be 0111 (for CH) and 0000 (for IB). MSB = 0 forces the Control Resistor to act as a resistor, reducing the voltage drop across the RRAM. Lower voltage drops lead to a longer spike time. After the spike corresponding to MSB = 0 is issued, the next MSB is either 1 or 0, depending on the desired spiking pattern.

## **7. Experimental Results:**

### *7.1 Experimental Setup*

The DC-IV measurements are performed using the Agilent B1500 semiconductor analyzer, while the transient measurements are performed using the B1530 Waveform Generator/Fast Measurement Unit (WGFMU). All the measurements are done at room temperature. To observe the spiking patterns in PMO RRAM device, waveforms of the voltage signals across the RRAM device from simulations are approximated and applied to the RRAM device during experiments.

### *7.2 Results and Discussion*

#### a. *Voltage Controlled Spiking Frequency*

The response of the neuron with a pulsed input voltage is presented in Fig. 12. The current response is shown for two different voltages $V_{input}$ = -1.6V, and $V_{input}$ = -1.8V, and it can be observed that the higher voltage led to faster spiking and vice versa. Fig. 12.c shows numerical of spiking frequency vs input voltage between the simulations and experiments. The average error is 4.4%.

#### b. *Time-Varying Input*

The same voltage signal as that in the simulation of the time-varying input signal (Fig. 10) is applied to the RRAM device. Voltage is reduced to 0 wherever necessary to mimic the resetting behavior. The results are presented in Fig. 13. The current response of the neuron shows dense spiking for high voltages, sparse spiking for moderate, and no spiking for low voltages. The experimental results corroborate the simulation results (Fig. 10).

#### c. *Different Spiking Patterns*

The CH and IB patterns are demonstrated experimentally and the results are presented in Fig. 14. To observe CH behaviour, three consecutive pulses of V = -2.4 and a single pulse of V = -1.7V are applied alternatively to the RRAM. To observe IB behavior, three consecutive pulses of V = -2.4V are applied to the RRAM, followed by consecutive pulses of V = -1.9V. The spiking patterns obtained experimentally qualitatively align with the simulation results (Fig. 11).

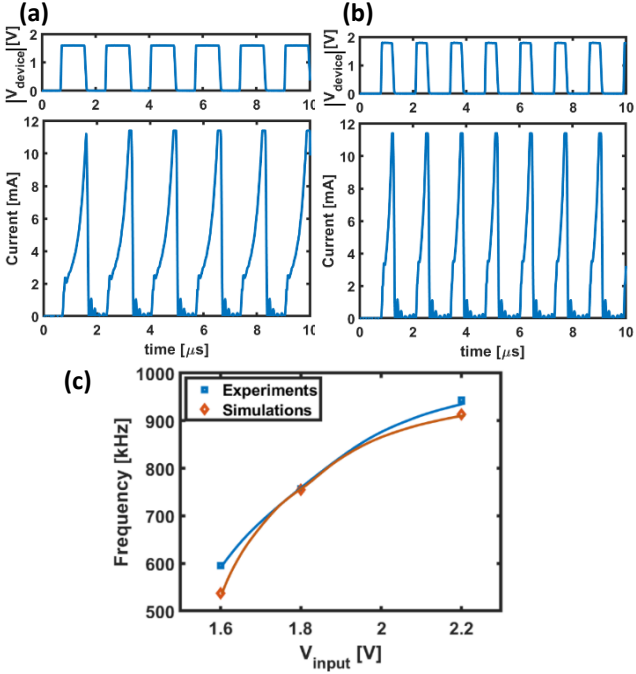

**Figure 12** Experimental Voltage Controlled Spiking Frequency behaviour demonstrated. For (a) $V_{input}$ = -1.6V, spiking frequency is 595 KHz while for (b) $V_{input}$ = -1.8V, spiking frequency is 757 KHz. (c) Comparison of Spiking Frequency vs Input voltage between experiments and simulations

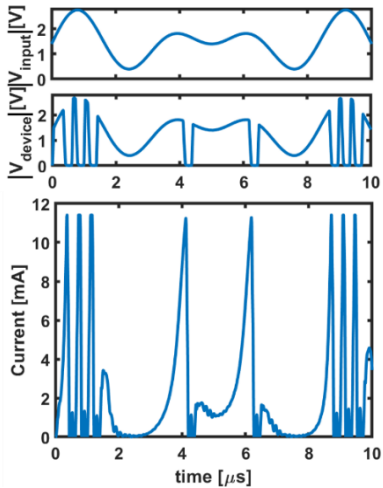

**Figure 13** Experimental Spiking Pattern for Time Varying Input: Input voltage is combination of two-time varying sinusoidal signals of equal magnitude and different frequencies.

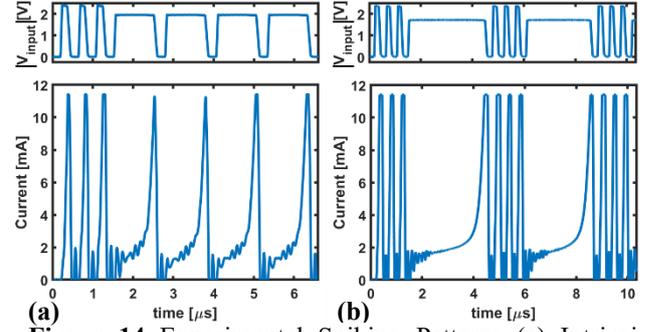

**Figure 14** Experimental Spiking Patterns (a) Intrinsic Bursting (b) Chattering

## 8. Relative Compactness of electrothermal vs. RC based timescale implementation

Fig. 15 shows two different implementations of a neuron using RRAM. Previously, RRAM has been used as a switching element only [19]-[20] (Fig. 15a). It has two resistance states with which the electrical capacitor either charges or discharges. In such neurons, the capacitor performs integration with an electrical timescale. The RRAM switching has its own timescale, which depends on the switching mechanism, e.g., IMT or self-heating. The relatively longer electrothermal RRAM switching timescales are proposed in this paper to replace the large electrical capacitors for integration functionality.

The electrothermal timescale from the 10μm ×10μm RRAM is experimentally shown as 100ns-1μs with a max current of 10mA. To implement an electrical RC timescale, assuming a 2nm thick SiO$_2$ based capacitor of the same area will produce capacitance of 1.7pF, which results in a time constant of 0.1ns for a 1V threshold for firing. This is a 100-1000x smaller timescale. For an equivalent timescale of 100ns-1μs, a 100-1000x larger area capacitor is needed with the same RRAM size to adversely affect area efficiency. As devices scale, the electrical-time constant is largely area independent – given by the following:

$$\tau_{RC} = \frac{CV}{I} = \left(\frac{\epsilon A}{d}\right)\frac{V}{\{J_D A\}} = \frac{\epsilon V}{d J_D} \quad (5)$$

Where **C** is electrical capacitance which depends on dielectric constant **ε**, the thickness of insulator **d**, and capacitor area **A**, **V** is the threshold voltage, **I** is charging current through the RRAM, which depends on the switching current density $J_D$ and area **A** assumed to be the same as the capacitor. The electrothermal timescale is also approximately scaling independent and is given by the following:

$$\tau_{th} = \frac{C_{th}\Delta T}{H} = C_v AL \frac{\Delta T}{\{V J_D AL\}} = (C_v) \frac{\Delta T}{\{V J_D\}} \quad (6)$$

Where $C_{th}$ is thermal capacitance, which depends upon specific heat capacity $C_v$, area of device A, and the thickness of the device, L, $\Delta T$ is a change in device temperature. So, the ratio of the timescale benefit will remain approximately constant. This largely sustains the large area efficiency with scaling.

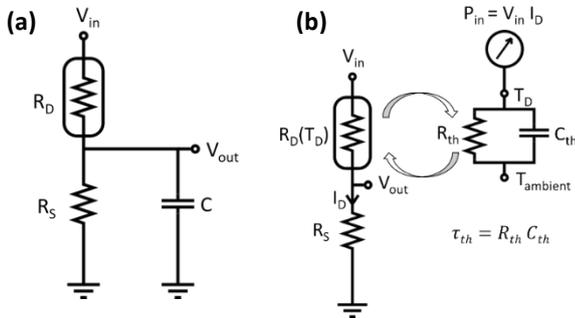

**Figure 15** (a) RRAM based neuron where RRAM device acts as a switching element and integration timescale is controlled by the external capacitor, (b) Proposed RRAM based neuron where RRAM device provides the electrothermal integration timescale in addition to being a switch.

## 9. Conclusion

In summary, we demonstrate different spiking patterns of a cortical neuron using a PMO RRAM-based neuron. The benchmarking of the PMO-based neuron circuit is shown in Table 2. In the proposed circuit, 6-bit register consisting of 6 flip flops, 1 OR gate, 2 switches, and 3 transistors for 1 resistor will be used, with a total transistor count of 119, based on which a feature size of 11.9 x$10^3$ $F^2$ is estimated. The neuron implementation shown in this paper is capacitor-less as PMO RRAM uses the internal self-heating timescales for integration operation and hence eliminates the use of a capacitor. Unlike previous demonstrations, an asynchronous simulation-based analysis with the driving circuitry and electrothermal model of the PMO RRAM device demonstrated realistic bursting patterns. Experiments guided by simulations validate the simulation results. With scalable PMO RRAM devices integrated with digital components for the driving circuit, a compact neuron can be designed, which is highly attractive for large-scale SNNs.

|  | Wijekoon [8] | Joubert [9] | Tuma [21] | Dutta [11] | Lashkare [13] | Lashkare [14] | Gao [19] | Shukla [20] | **This Work** |
|---|---|---|---|---|---|---|---|---|---|
| Platform | CMOS | CMOS | Phase Change + CMOS | SOI CMOS | PCMO + PMOS | PMO + CMOS | - | - | **PMO + CMOS** |
| Circuit Type | Analog + Asynch. | Digital + Synch. | Mixed + Synch. | Analog + Asynch. | Mixed + Synch. | Mixed + Asynch. | - | - | **Mixed + Asynch.** |
| Neuron Model | LIF | LIF | IF | LIF | IF | LIF | LIF | LIF | **LIF** |
| Spiking Behavior* | RS, FS, IB, CH | RS | RS | RS | RS | RS | - | - | **RS, IB, CH** |
| Refractory Period | Fixed | Fixed | Fixed | Fixed | Fixed | Fixed | - | - | **Control** |
| Timescale Generation | Cap | Cap | - | Floating Body Effect | Gradual Resistive Switching | Electro-thermal | Cap | Cap | **Electro-thermal** |
| RRAM Usage | - | - | - | - | Integrator | Integrator | Switch | Switch | **Integrator, Refractory Period** |
| Area (×$10^3$ $F^2$) | 23 | 127 | 2551 | 1.767 | 3.086 | - | - | - | **11.908** |

**Table 2:** Benchmarking with Previous RRAM Implementations
*RS: Regular Spiking, FS: Fast Spiking

**References**

[1] Izhikevich, E.M., 2004. Which model to use for cortical spiking neurons?. *IEEE transactions on neural networks*, *15*(5), pp.1063-1070.

[2] Santurkar, S. and Rajendran, B., 2015, July. C. elegans chemotaxis inspired neuromorphic circuit for contour tracking and obstacle avoidance. In *International Joint Conference on Neural Networks (IJCNN)* (pp. 1-8).

[3] Weistuch, C., Mujica-Parodi, L.R. and Dill, K., 2021. The refractory period matters: unifying mechanisms of macroscopic brain waves. *Neural Computation*, *33*(5), pp.1145-1163.

[4] Santurkar, S. and Rajendran, B., 2015, July. C. elegans chemotaxis inspired neuromorphic circuit for contour tracking and obstacle avoidance. In *2015 International Joint Conference on Neural Networks (IJCNN)* (pp. 1-8). IEEE.

[5] Izhikevich, E.M., 2003. Simple model of spiking neurons. *IEEE Transactions on neural networks*, 14(6), pp.1569-1572.

[6] Babacan, Y., Kaçar, F. and Gürkan, K., 2016. A spiking and bursting neuron circuit based on memristor. *Neurocomputing*, *203*, pp.86-91.

[7] Sourikopoulos, I., Hedayat, S., Loyez, C., Danneville, F., Hoel, V., Mercier, E. and Cappy, A., 2017. A 4-fJ/spike artificial neuron in 65 nm CMOS technology. *Frontiers in neuroscience*, *11*, p.123.

[8] Wijekoon, J.H. and Dudek, P., 2008. Compact silicon neuron circuit with spiking and bursting behaviour. *Neural Networks*, *21*(2-3), pp.524-534.

[9] Joubert, A., Belhadj, B. and Héliot, R., 2011, June. A robust and compact 65 nm LIF analog neuron for computational purposes. In *IEEE 9th International New Circuits and systems conference* (pp. 9-12).

[10] Hynna, K.M. and Boahen, K., 2007, May. Silicon neurons that burst when primed. In *IEEE International Symposium on Circuits and Systems* (pp. 3363-3366).

[11] Dutta, S., Kumar, V., Shukla, A., Mohapatra, N.R. and Ganguly, U., 2017. Leaky integrate and fire neuron by charge-discharge dynamics in floating-body MOSFET. *Scientific reports*, *7*(1), pp.1-7.

[12] Soleimani, H., Ahmadi, A. and Bavandpour, M., 2012. Biologically inspired spiking neurons: Piecewise linear models and digital implementation. *IEEE Transactions on Circuits and Systems I: Regular Papers*, *59*(12), pp.2991-3004.

[13] Lashkare, S., Chouhan, S., Chavan, T., Bhat, A., Kumbhare, P. and Ganguly, U., 2018. PCMO RRAM for integrate-and-fire neuron in spiking neural networks. *IEEE Electron Device Letters*, *39*(4), pp.484-487.

[14] Lashkare, S., Bhat, A., Kumbhare, P. and Ganguly, U., 2018, October. Transient joule heating in PrMnO3 RRAM enables ReLU type neuron. In *Non-Volatile Memory Technology Symposium (NVMTS)* (pp. 1-4).

[15] Chavan, T., Dutta, S., Mohapatra, N.R. and Ganguly, U., 2020. Band-to-band tunneling based ultra-energy-efficient silicon neuron. *IEEE Transactions on Electron Devices*, *67*(6), pp.2614-2620.

[16] Chakraborty, I., Singh, A.K., Kumbhare, P., Panwar, N. and Ganguly, U., 2015, June. Materials parameter extraction using analytical models in PCMO based RRAM. In *73rd Annual Device Research Conference (DRC)* (pp. 87-88).

[17] Lampert, M.A., 1964. Volume-controlled current injection in insulators. *Reports on Progress in Physics*, *27*(1), p.329.

[18] Cong, B.T., Tsuji, T., Thao, P.X., Thanh, P.Q. and Yamamura, Y., 2004. High-temperature thermoelectric properties of Ca1−xPrxMnO3−δ (0⩽ x< 1). *Physica B: Condensed Matter*, *352*(1-4), pp.18-23.

[19] Gao, L., Chen, P.Y. and Yu, S., 2017. NbOx based oscillation neuron for neuromorphic computing. *Applied physics letters*, *111*(10), p.103503.

[20] Shukla, N., Parihar, A., Freeman, E., Paik, H., Stone, G., Narayanan, V., Wen, H., Cai, Z., Gopalan, V., Engel-Herbert, R. and Schlom, D.G., 2014. Synchronized charge oscillations in correlated electron systems. *Scientific reports*, *4*(1), pp.1-6.

[21] Tuma, T., Pantazi, A., Le Gallo, M. *et al.* Stochastic phase-change neurons. *Nature Nanotech* **11,** 693–699 (2016). https://doi.org/10.1038/nnano.2016.70